\documentclass[aps,prx,reprint,superscriptaddress,amsmath,amssymb]{revtex4-2}

\usepackage{graphicx}
\usepackage[driverfallback=dvipdfm]{hyperref}
\usepackage{tabularx}
\usepackage{mathtools}
\usepackage[mediumspace,mediumqspace,squaren]{SIunits}
\usepackage{chngcntr}

\usepackage{xcolor}

\newcommand{\ee} {{\textrm e}}
\newcommand{\dd} {\hbox{\textrm d}}

\newcommand{\erfc} {\hbox{\textrm erfc}}
\newcommand{\Real}{\text{Re}}
\newcommand{\Imag}{\text{Im}}

\newcommand{\bra}[1]{\ensuremath{\left\langle #1 \right|}}
\newcommand{\ket}[1]{\ensuremath{\left| #1 \right\rangle}}
\newcommand{\braket}[2]{\left\langle #1\middle|#2\right\rangle}
\newcommand{\ketbra}[2]{\left| #1 \middle\rangle\middle\langle #2 \right|}

\newcommand{\os} {{\hat{\sigma}}}
\newcommand{\oG} {{\hat{G}}}
\newcommand{\oH} {{\hat{H}}}

\newcommand{\oP} {{\hat{P}}}
\newcommand{\oQ} {{\hat{Q}}}
\newcommand{\oU} {{\hat{U}}}
\newcommand{\oD} {{\hat{D}}}
\newcommand{\ox} {{\hat{x}}}
\newcommand{\op} {{\hat{p}}}

\newcommand{\orho} {{\hat{\rho}}}

\begin{document}
\title{Protecting collective qubits from non-Markovian dephasing}

\author{Antoine Covolo} \affiliation{JEIP,  UAR 3573 CNRS, Coll{\`e}ge de France, PSL University, 11, place Marcelin Berthelot, 75231 Paris Cedex 05, France}

\author{Valentin Magro} \affiliation{JEIP,  UAR 3573 CNRS, Coll{\`e}ge de France, PSL University, 11, place Marcelin Berthelot, 75231 Paris Cedex 05, France}

\author{Mathieu Girard} \affiliation{JEIP,  UAR 3573 CNRS, Coll{\`e}ge de France, PSL University, 11, place Marcelin Berthelot, 75231 Paris Cedex 05, France}

\author{S{\'e}bastien Garcia} \affiliation{JEIP,  UAR 3573 CNRS, Coll{\`e}ge de France, PSL University, 11, place Marcelin Berthelot, 75231 Paris Cedex 05, France}

\author{Alexei Ourjoumtsev} \email{Corresponding author: alexei.ourjoumtsev@college-de-france.fr} \affiliation{JEIP,  UAR 3573 CNRS, Coll{\`e}ge de France, PSL University, 11, place Marcelin Berthelot, 75231 Paris Cedex 05, France}

\begin{abstract} 

In quantum technologies, qubits encoded as collective atomic or solid-state excitations present important practical advantages, such as strong directional coupling to light. Unfortunately, they are affected by inhomogeneities between the emitters, which make them decay into dark states. In most cases, this process is non-Markovian. Through a simple formalism, we unveil a regime where this decay is suppressed by a combination of driving and non-Markovianity. We experimentally demonstrate this ``driving protection'' using a Rydberg superatom, making its coherent Rabi oscillations last about $14$ times longer than the free lifetime of the collective qubit.

\end{abstract}

\maketitle

\section{Introduction}
Quantum systems are never fully closed. Describing their interactions with their classical environment in an accurate and tractable way is essential in various contexts, in particular in quantum technologies. The textbook scenario assumes that these interactions are constant, and that the resulting correlations between the system and the environment instantly vanish without affecting the latter. For a qubit carried by a two-level atom, these interactions lead to three effects: the decay of the excited state, its excitation by thermal photons, and pure dephasing. In this Born-Markov approximation, the coherence of the system always decays exponentially.

Qubits can also be encoded as collective excitations of many emitters, acting together like a quantum phased array antenna. The excited state of the qubit is then an exchange-symmetric Dicke state \cite{Dicke1954}, with a single excitation delocalized over the whole ensemble. This encoding is the cornerstone of heralded quantum information processing with atomic ensembles \cite{Duan2001}, of quantum memories using rare-earth ion-doped crystals or atomic gases \cite{Lvovsky2009}, and of recent quantum optics experiments using Rydberg superatoms \cite{Pedersen2008,Kumlin2023,Shao2024}. Among other practical advantages, it offers a  directional coupling to light enhanced with respect to that of a single emitter \cite{Dudin2012b}. However, collective encoding suffers from an inherent drawback: crystalline defects, inhomogeneous external fields, or thermal atomic motion randomly shift the transition frequencies of the emitters. With time, the phases between the elements of the ``antenna'' become random, scrambling its emission. In other words, inhomogeneous dephasing makes the qubit's excited state decay in a quasi-continuum of dark, asymmetric Dicke states.

The Born-Markov approximation generally fails to describe this effect. To see this, let us consider 
$N\gg 1$ two-level emitters with angular frequencies $\omega_n$, carrying a qubit encoded in the states $\ket{G}=\prod_{n=1}^N|g^{(n)}\rangle$ where all the emitters are in the ground state $g$, and $\ket{E_0}=\sum_{n=1}^N\os_{eg}^{(n)}\ket{G}/\sqrt{N}$ where $\os_{eg}^{(n)}=|e^{(n)}\rangle\langle g^{(n)}|$ promotes the emitter $n$ into the excited state $e$, such that the excitation is fully delocalized. After a time $t$, each term in this sum accumulates a phase factor $\ee^{-i\omega_n t}$. If the frequencies $\omega_n$ follow a probability distribution $P(\omega)$, an ergodic approximation shows that the overlap between the initial state $\ket{E_0}$ and the final state \ket{E(t)} obeys
\begin{equation}
\label{EqDecayEFT}
s(t)=\overline{\braket{E_0}{E(t)}}=\frac{1}{N}\sum_{n=1}^N\overline{\ee^{-i\omega_n t}} = \int_{-\infty}^{+\infty}\dd\omega P(\omega)\ee^{-i\omega t}.
\end{equation}
Unless $P(\omega)$ is a Lorentzian, the decay of $s(t)$ is not exponential: the quasi-continuum of dark states has a memory, allowing some information to flow back into the system.

Equation \ref{EqDecayEFT} is sometimes used as-is \cite{Zhao2009,SchmidtEberle2020}. However, it assumes that the system is excited at a well-known moment, quasi-instantaneously with respect to the dephasing. While reasonable for heralded collective atomic spin qubits \cite{Duan2001}, this assumption fails for temporally multimode quantum memories \cite{Nunn2008}, where the preparation time is undefined, and for Rydberg superatoms, where the dephasing is currently not negligible at the timescales of single-qubit gates. Moreover, as shown below, this decay can dramatically change in presence of driving. One is then left with two unappealing choices: keeping $N-1$ irrelevant states in the density matrix, or using a generalized master equation \cite{Breuer2007} to describe the evolution. Both are resource-intensive, and obscure much of the relevant physics.

\begin{figure}[tb]
\centering
\includegraphics[width=85mm]{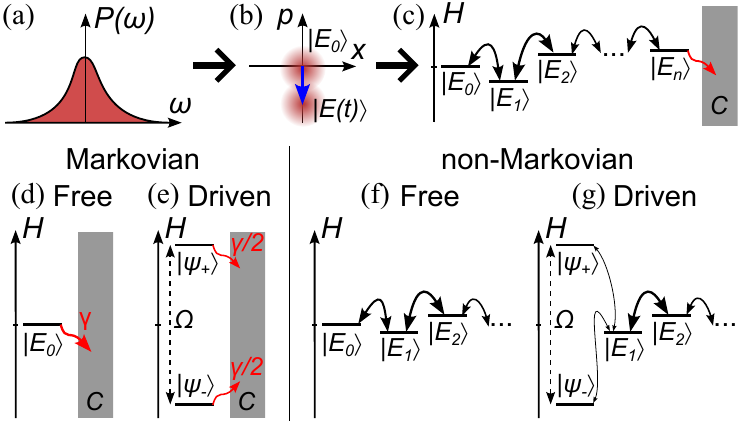}
\caption{ \textbf{Modeling inhomogeneous dephasing and driving protection.} The effect of an inhomogeneous dephasing spectrum $P(\omega)$ (a) is mapped to a displacement in a phase space where frequency and time play the roles of the usual position and momentum quadratures $x$ and $p$ (b). This allows one to derive the corresponding ladder of $n+1$ dephasing-coupled Dicke states \ket{E_{k\leq n}} and their coupling coefficients (c), included in the Hamiltonian to describe the non-Markovian features, and to determine the irreversible decay of \ket{E_n} into the ``continuum'' $C=\{\ket{E_{k>n}}\}$. In the Markovian case, \ket{E_0} decays directly in the continuum (d). When driving the qubit transition with a Rabi frequency $\Omega$ (e), the dressed eigenstates $\ket{\psi_\pm}=(\ket{G}\pm\ket{E_0})/\sqrt{2}$ also decay in the continuum, owing to its broad spectral extent. In a non-Markovian case, the decay out of the qubit's subspace decreases drastically between the free dephasing from \ket{E_0} (f) and the driven case (g), due to the energy difference between \ket{\psi_\pm} and the \ket{E_{k\geq 1}} asymmetric Dicke states.  }
\label{fig:DephDisplDicke}
\end{figure}

Here, we show that the non-Markovianity of this decay process can not only be efficiently modeled, but also used to suppress it with remarkably little experimental effort. By mapping the inhomogeneous dephasing onto a displacement in time-frequency phase space, we construct a ladder of sequentially coupled Dicke states (Fig.~\ref{fig:DephDisplDicke}(a),(b),(c)) and show that they can be determined analytically in many physically-relevant cases. Including the first few of them in the system's Hilbert space and treating the others as a broad continuum provides a precise and resource-efficient description of the non-Markovian dynamics. In an analytically-tractable limit, we identify a process akin to cavity protection \cite{Diniz2011,Kurucz2011,Baghdad2023,Sauerwein2023} 
where the qubit's decay is inhibited by a combination of non-Markovianity and strong driving (Fig.~\ref{fig:DephDisplDicke}(d-g)). We experimentally demonstrate this effect by using a Rydberg superatom undergoing two spectrally different non-Markovian processes. Consistently with the predictions of our model, we observe that its coherent Rabi oscillations can last up to $\sim 14$ times longer than the characteristic inhomogeneous dephasing time.

\section{Theoretical model}

\label{sec:Model}

Using the definitions above, let us consider the effect of the inhomogeneous dephasing Hamiltonian $\oH=\sum_{n=1}^N\omega_n\os_{ee}^{(n)}$ on a single excitation in the ensemble ($\hbar=1$ in the following). We rewrite $\omega_n=\sqrt{2} \omega_{\mathrm{S}}\,x_n$ where $x_n$ is a dimensionless random variable governed by a probability distribution $\rho(x)=\sqrt{2} \omega_{\mathrm{S}} P(\sqrt{2}\omega_{\mathrm{S}} \, x)$. The scaling rate $\omega_{\mathrm{S}}$ leaves some freedom for optimizing the latter. 
Assuming ergodicity with $N\gg 1$, we approximate the sum over $n$ by $\int_{-\infty}^{+\infty}\dd x$ written as $\int \dd x$ to simplify notations, and replace the discrete operators $\os_{ij}^{(n)}$ with $\os_{ij}(x)$ satisfying
$
[\os_{ij}(x),\os_{kl}(x')]=\delta(x-x')(\delta_{jk}\os_{il}(x)-\delta_{il}\os_{kj}(x)).
$ 
We then rewrite $\oH=\sqrt{2} \omega_{\mathrm{S}} \int\dd x\, x \,\os_{ee}(x)$ and $\ket{E_0}=\int\dd x \, \psi_0(x) \os_{eg}(x)\ket{G}$ where the wavefunction $\psi_0(x)=\sqrt{\rho(x)}\ee^{i\phi(x)}$ satisfies $|\psi_0(x)|^2  = \rho(x)$, the arbitrary phase $\phi(x)$  leaving the choice of the most convenient gauge.

In the single excitation regime considered here, the dimensionless frequency $x$ and its conjugate dimensionless time $p$ can be treated as operators equivalent to position and momentum, with $[\ox,\op]=i$ \cite{Fabre2022}. Then $\os_{eg}(x)\ket{G}=\ket{x}$ is formally equivalent to a usual continuous-variable position eigenstate, giving $\ket{E_0}=\int\dd x\, \psi_0(x) \ket{x}$ and $\oH=\sqrt{2} \omega_{\mathrm{S}} \int\dd x\, x \ketbra{x}{x}=\sqrt{2}\omega_{\mathrm{S}}\, \ox$, which allows us to interpret the evolution $\hat{U}(t)=\exp(-i\oH t)$ as a phase-space displacement in the direction $-p$ by a distance $\sqrt{2}\omega_{\mathrm{S}}\, t$. Eq.~\ref{EqDecayEFT} is recovered via \begin{equation}
\label{EqDecaySvsX}
s(t)=\bra{E_0}\hat{U}(t)\ket{E_0}=\int\dd x \rho(x)\ee^{-i\sqrt{2}\omega_{\mathrm{S}} \, x \, t}.
\end{equation}

We consider three physical examples: 
\begin{itemize}
\item A Markovian process, with a Lorentzian spectrum $\rho_{\mathrm{M}}(x) = 2 / ( \pi (1 + 4 x^2))$ and a population decay rate $\sqrt{2}\omega_{\mathrm{M}}$, where we can choose $\psi_{\mathrm{M},0}(x) = \sqrt{2/\pi} / (1 + i 2 x)$. As expected, Eq.~\ref{EqDecaySvsX} leads to an exponential decay of the amplitude, $s_{\mathrm{M}}(t) = \ee^{-\omega_{\mathrm{M}} t /\sqrt{2} }$.
\item Gaussian inhomogeneous dephasing, with $\rho_{\mathrm{G}}(x)=\ee^{-x^2}/\sqrt{\pi}$ and $\psi_{G,0}(x) = \sqrt{\rho_{\mathrm{G}}(x)}$, encountered for example in atomic ensembles where the velocities $v_n$ obey Maxwell-Boltzmann statistics. The Doppler shifts $\omega_n=k\,v_n$ then follow a Gaussian distribution where the width $\omega_{\mathrm{G}}=k\sqrt{k_{\mathrm{B}} T/m}$ depends on the effective spin-wave vector $k$, the Boltzmann constant $k_{\mathrm{B}}$, the temperature $T$ and the atomic mass $m$. The state \ket{E_0}, which corresponds to a vacuum state \ket{0} in this phase space, is displaced by $\hat{U}(t)$ into a coherent state \ket{-i\omega_\mathrm{G} t}, with a Gaussian decay of the amplitude $s_{\mathrm{G}}(t)=\braket{0}{-i\omega_\mathrm{G} t}=\ee^{-\omega_{\mathrm{G}}^2t^2/2}$. 
\item Differential light shifts between ground and excited atomic states in a three-dimensional harmonic trap, where the Gaussian thermal distribution of atomic positions and the quadratic position-dependance of the light-shift potential lead to $\rho_{\mathrm{L}}(x)=\sqrt{x} \ee^{-x} \theta(x) 2/\sqrt{\pi}$ and $\psi_{L,0}(x) = \sqrt{\rho_{\mathrm{L}}(x)}$, where $\theta$ is the Heaviside step function. The scaling rate is $\omega_{\mathrm{L}} = \delta\alpha_{\mathrm{r}} \, k_{\mathrm{B}} T / \sqrt{2} \hbar $ where $\delta\alpha_{\mathrm{r}} = (\alpha_e - \alpha_g) / \alpha_g $ is the relative difference of polarizabilities of the excited and ground states. The amplitude $s$ decays here as $s_{\mathrm{L}}(t) =(1 + i \sqrt{2} \omega_\mathrm{L} t )^{-3/2}$.
\end{itemize}
Unless stated otherwise, we set $\omega_{\mathrm{S}}=1$ as a unit for angular frequencies and rates, and $1/\omega_{\mathrm{S}}$ as a unit for time.

\medskip
The interest of this mapping becomes apparent when the qubit experiences classical or quantum driving and Eqs.~\ref{EqDecayEFT} or~\ref{EqDecaySvsX} no longer hold. As shown below, it allows one to explicitly construct the ladder of Dicke states sequentially coupled by the inhomogeneous dephasing. By truncating this ladder at a given level and treating its coupling with higher-order states as an irreversible decay, one can use a standard  Gorini - Kossakowski - Sudarshan - Lindblad (GKSL) master equation \cite{Gorini1976,Lindblad1976} to describe the non-Markovian dynamics via a small extension of the Hilbert space.

We start by constructing the ladder of Dicke states $\{\ket{E_k}\}$ from $\ket{E_0}$ by successive applications of $\oH$. As $\oH\propto\hat{x}$, the corresponding wavefunctions have the form $\psi_k(x)=Q_k(x)\psi_0(x)$ where $Q_k$ is a polynomial of degree $k$. The problem then reduces to finding a family of orthogonal polynomials for the weight function $\rho$: $\langle E_k|E_l \rangle=\int \rho(x) Q_k^*(x)Q_l(x)=\delta_{k,l}$.
This problem is generically solved by Gram-Schmidt orthonormalization but, given its importance in mathematics, engineering or data science, explicit solutions are known in many cases~\cite{Szego1975}. By construction, $\oH$ is tridiagonal in this basis, with nonzero elements $H_k=\bra{E_k}\oH\ket{E_k}$ and $V_k=\bra{E_{k-1}}\oH\ket{E_k}$. For the Gaussian $\rho_\mathrm{G}$, $Q_k$ is a Hermite polynomial up to a normalization factor and \ket{E_{\mathrm{G},k}} is the Fock state \ket{k}, which directly gives $H_{\mathrm{G},k}=0$ and $V_{\mathrm{G},k}=\sqrt{k}$. For the light shift $\rho_\mathrm{L}$, $Q_k(x)=\sqrt{2^k k!/(2k+1)!!}L_k^{(1/2)}(x)$
where $L_k^{(1/2)}$ is a generalized Laguerre polynomial. Their recurrence relation~\cite{Szego1975} gives $H_{\mathrm{L},k}=\sqrt{2}(2k+3/2)$ and $V_{\mathrm{L},k}=-\sqrt{k(2k+1)}$. For a generic $\rho(x)$ we initialize the recurrence with $\psi_{-1}(x)=0$ and $\psi_0(x)$, calculate
$a_k\psi_{k+1}(x)=(\sqrt{2}x-H_k)\psi_{k}(x)-V_k\psi_{k-1}(x)$,
and determine $a_k$ by normalization.

If $\rho(x)\sim x^{-d}$ at infinity, the ladder of Dicke states terminates at step $n=\lfloor (d-1)/2\rfloor$ as $x\,\psi_n(x)$ becomes non-normalizable: the effect of $\oH$ on \ket{E_n} is then treated in a Born-Markov approximation as an irreversible decay into a spectrally-broad continuum. For the Lorentzian $\rho_\mathrm{M}$ this happens at $n=0$. If this ``quantum Markov order'' \cite{Taranto2019a} is impractically large or infinite, as it is the case for $\rho_\mathrm{G}$ and $\rho_\mathrm{L}$, one can choose $n$ as a compromise between the precision and the size of the simulations and approximate the coupling of \ket{E_n} to higher-lying states by a Markovian process. 

\begin{figure}[tb]
\centering
\includegraphics[width=85mm]{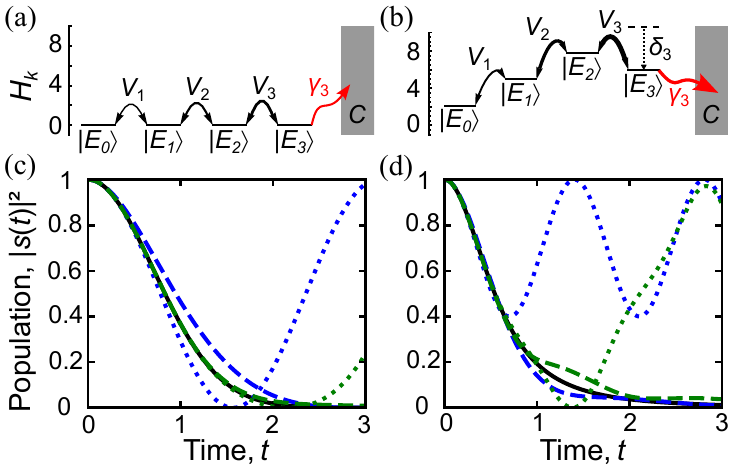}
\caption{ \textbf{Dicke basis truncation.} (a) and (b) Energy levels $H_k$ of truncated basis states with $n=3$ for the Gaussian and lighshift-broadened cases, respectively. Arrow linewidths are proportional to coupling coefficients $|V_k|$ or decay rates $\gamma_3$ towards the continuum \ket{C}. The dashed line for $E_3$ represent the energy $H_3$ before the addition of the Lamb-like energy shift $\delta_3$. (c) and (d) Related evolutions of the population $|s(t)|^2$. The black curves are the analytic solutions given by Eq.~\ref{EqDecaySvsX}. The blue dotted and dashed lines are computed in a truncated basis with $n=1$, without and with the continuum state, respectively. The green curves use the basis with $n=3$, again without or with the continuum.  }
\label{fig:TB}
\end{figure}

To terminate this ladder, we thus need to calculate the amplitude decay rate $\gamma_n$ and the Lamb-like energy shift $\delta_n$ due to the coupling of \ket{E_n} with the ``broad continuum'' of states 
$\{\ket{E_{k>n}}\}$. A straightforward use of the resolvent $\oG(z)=(z-\oH)^{-1}$ \cite{Cohen1998.ch3}, detailed in Appendix~\ref{app:decay}, shows that $\Delta_n=\delta_n-i\gamma_n$ obeys the recurrence relation
\begin{align}
\label{eq:DeltaRec}
\Delta_n =& H_0-H_n-\frac{|V_n|^2}{\Delta_{n-1}},
\end{align}
the initial value $\Delta_0$ being given by $\Delta_0=-1/G_0(H_0)$ where, in the limit $\Imag(z)\rightarrow 0^+$,
\begin{multline}
G_0(z)
= \bra{E_0}\oG(z)\ket{E_0}\\
= \int \frac{\rho(x)\dd x}{z-\sqrt{2} x}
= -i \int_0^{+\infty}\ee^{izt}s(t)\dd t
\end{multline}
For the Markovian case $\rho_\mathrm{M}$, $G_{\mathrm{M},0}(z)=1/(z+i/\sqrt{2})$ and $H_{\mathrm{M},0}=0$ gives, self-consistently, $\delta_{\mathrm{M},0}=0$ and $\gamma_{\mathrm{M},0}=1/\sqrt{2}$.
For the Gaussian $\rho_\mathrm{G}$ with $H_{\mathrm{G},0}=0$ and 
$G_{\mathrm{G},0}(z) = -i\sqrt{\pi/2}\,w(z/\sqrt{2})$ where $w(z)=\ee^{-z^2}\erfc(-iz)$ is the Faddeeva function, $\delta_{\mathrm{G},n}=0$ and
 $\gamma_{\mathrm{G},n}=(2/\pi)^{(-1)^n/2}n!!/(n-1)!!$. For the light-shift spectrum $\rho_\mathrm{L}$, $G_{\mathrm{L},0}(z) = -\sqrt{2}-i\sqrt{\pi z} \sqrt[4]{2}\,w(\sqrt{z}/\sqrt[4]{2})$ and $H_{\mathrm{L},0}=3/\sqrt{2}$; Eq.~\ref{eq:DeltaRec} becomes $\Delta_{\mathrm{L},n}=-2^{3/2}n-n(2n+1)/\Delta_{\mathrm{L},n-1}$ and gives $\delta_{\mathrm{L},n}$ and $\gamma_{\mathrm{L},n}$ by induction.

As shown in Fig.~\ref{fig:TB}, numerical simulations in a basis of size $n+1$ rapidly converge to the analytic solutions of Eq.~\ref{EqDecaySvsX} with increasing $n$. At long times, approximating the higher-lying states by a Markovian continuum significantly improves the precision compared to a simple phase-space truncation. This approach enables resource-efficient numerical simulations in more complex situations where several collective qubits are coupled to time-dependent classical or quantum light.

\section{Driving protection effect}
\label{sec:DecoFree}

This formalism predicts an interesting effect when the qubit is resonantly driven with a constant Rabi angular frequency $\Omega$. A similar resolvent-based approach, detailed in Appendix~\ref{app:protection} together with the derivations of the expressions below, shows that Eq.~\ref{EqDecaySvsX}, describing the excited-state amplitude dynamics, is then replaced with
\begin{equation}
\label{eq:SOmega}
s_\Omega(t)
= \int_{-\infty}^{+\infty}\frac{i\dd z}{2\pi}\frac{zG_0(z+H_0)\ee^{-itz}}{z-\frac{\Omega^2}{4}G_0(z+H_0)}.
\end{equation}
In the following we focus on the strong-driving regime where $\Omega\gg 1$.

For the Lorentzian spectrum $\rho_\mathrm{M}$, $H_{\mathrm{M},0}=0$ and the simple form of $G_{\mathrm{M},0}(z)$ gives the exact solution
\begin{equation}
\label{eq:SMOmega}
s_{\mathrm{M},\Omega}(t)=\ee^{-\gamma_\mathrm{M}t/2}\left(\cos(\Omega_\mathrm{M}t/2) - \frac{\gamma_\mathrm{M}}{\Omega_\mathrm{M}}\sin(\Omega_\mathrm{M}t/2) \right)
\end{equation}
with $\Omega_\mathrm{M}=(\Omega^2-\gamma_\mathrm{M}^2)^{1/2}$. In this Markovian case, a conventional master equation approach leads to the same result. For $\Omega\geq \gamma_\mathrm{M}=1/\sqrt{2}$ the decay rate $\gamma_\mathrm{M}/2$, reduced by half by the driving, becomes independent on $\Omega$: no matter the driving strength, the decay persists.

\begin{figure}[tb]
\centering
\includegraphics[width=85mm]{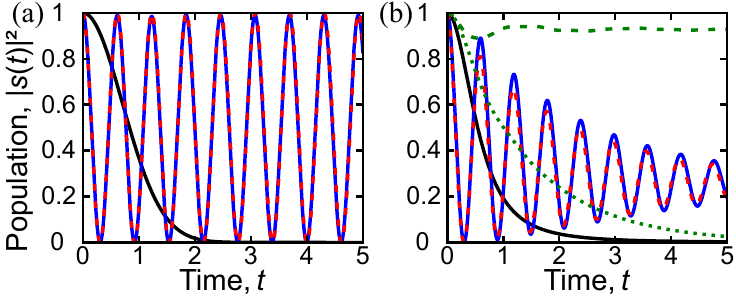}
\caption{ \textbf{Driving protection.} Evolution of the population $|s_\Omega(t)|^2$ (a) for the Gaussian spectrum  $\rho_\mathrm{G}$ and (b) for the lightshift-broadened spectrum $\rho_\mathrm{L}$, with and without driving. The black curves are the analytic solutions of Eq.~\ref{EqDecaySvsX} for an undriven qubit ($\Omega=0$). The blue lines are computed in a Dicke basis with a Rabi frequency $\Omega=10$. The dotted red curves are the corresponding approximations given by Eqs.~\ref{eq:SGOmegaAppr} and~\ref{eq:SLOmegaAppr}. In (b), the dotted and dashed green curves are the populations of \ket{\psi_+} and \ket{\psi_-} states, respectively.}
\label{fig:ThDriveProtect}
\end{figure}

The situation radically changes for the Gaussian spectrum $\rho_\mathrm{G}$. The asymptotic expansion $\Real(G_{\mathrm{G},0}(z))=z^{-1}+z^{-3}+O(z^{-5})$ gives
\begin{align}
\nonumber
\mathrlap{s_{\mathrm{G},\Omega}(t)
\approx 
\ee^{-\gamma_{\mathrm{G},\Omega} t}\cos(\Omega_\mathrm{G}t/2),}\\
\label{eq:SGOmegaAppr}
\Omega_\mathrm{G}
\approx &~ \Omega+\frac{2}{\Omega},
&
\gamma_{\mathrm{G},\Omega}
\approx &~
\frac{\Omega^2}{8}\sqrt{\frac{\pi}{2\ee}}\ee^{-\Omega^2/8} .
\end{align}
The decay profile changes from Gaussian to exponential, its rate $\gamma_{\mathrm{G},\Omega}$ vanishing exponentially with $\Omega^2$. Here, strong driving combined with non-Markovianity allows high-visibility Rabi oscillations to occur over times strongly exceeding the natural dephasing time, as shown in Fig.~\ref{fig:ThDriveProtect}(a). This ``driving protection'' of the collective qubit is closely related to cavity protection \cite{Diniz2011,Kurucz2011,Baghdad2023,Sauerwein2023}, the collective Rabi frequency $\Omega$ replacing the collective cavity coupling.

For the light-shift spectrum $\rho_\mathrm{L}$, the expansion $\Real(G_{\mathrm{L},0}(z+H_{\mathrm{L},0}))=z^{-1}+3z^{-3}+O(z^{-4})$ leads to
\begin{align}
\nonumber
\mathrlap{s_{\mathrm{L},\Omega}(t)
\approx 
\frac{1}{2}\left(\ee^{-\gamma_{\mathrm{L},\Omega} t}\ee^{-i\Omega_\mathrm{L} t/2}+\ee^{i\Omega_\mathrm{L} t/2}\right),}\\
\label{eq:SLOmegaAppr}
\Omega_\mathrm{L}
\approx &~ \Omega+\frac{6}{\Omega},
&
\gamma_{\mathrm{L},\Omega}
\approx &~
\frac{\Omega^2}{8}\sqrt{\frac{\pi\Omega}{\sqrt{2}\ee^3}}\ee^{-\Omega/\sqrt{8}} .
\end{align}
As visible in Fig.~\ref{fig:ThDriveProtect}(b), the driving protection persists but, since $\gamma_{\mathrm{L},\Omega}$ decreases slower than $\gamma_{\mathrm{G},\Omega}$ with $\Omega$, it is weaker than for a Gaussian spectrum. While in the Gaussian case the system fully decays in the continuum, here the population $|s_{\mathrm{L},\Omega}(t)|^2$ tends to $1/4$ at long times. In fact, the two terms in the expression of $s_{\mathrm{L},\Omega}(t)$ respectively correspond to the propagators $\bra{\psi_\pm}\oU_\Omega(t)\ket{\psi_\pm}$ with $\ket{\psi_\pm}=(\ket{G}\pm\ket{E_0})/\sqrt{2}$: the population in \ket{\psi_+} decays with a rate $2\gamma_{\mathrm{L},\Omega}$ while \ket{\psi_-} is very efficiently protected by the drive.

Figure~\ref{fig:ThDriveProtect} shows that the approximate equations \ref{eq:SGOmegaAppr} and~\ref{eq:SLOmegaAppr} correctly reproduce the dynamics computed in sufficiently large Dicke bases. For the light-shift spectrum, the numerical integration confirms that \ket{\psi_-} is efficiently protected. Poles of the integrand in Eq.~\ref{eq:SOmega} neglected in the approximations above account for transient dynamics at $0<t\lesssim 1$.

\section{Physical interpretation}

While this ``driving protection'' qualifies as dynamical decoupling, it is a fundamentally collective effect, different from the techniques generally used to counteract dephasing in two-level systems. The simplest of them is spin echo \cite{Hahn1950}, where the dephasing due to an unknown shift $\omega$ of the resonance frequency can be compensated by periodically swapping the ground and excited amplitudes via $\pi$ pulses. This mitigation persists when the qubit is driven continuously with a Rabi frequency $\Omega_0\gg\omega$ \cite{Viola1998}. 

The process considered here is not a mere sum of such effects acting on every emitter in the ensemble. Indeed, the Rabi frequency $\Omega=\Omega_0\sqrt{N}$ is collectively enhanced with respect to the individual Rabi frequency $\Omega_0$. The latter can be much smaller than $\omega_\mathrm{S}$, and thus insufficient to mitigate dephasing in a single emitter. Moreover, discrete or continuous spin echo fundamentally relies on swapping the roles of the ground and the excited states. Here the excited population of each emitter remains below $1/N$. For Rydberg superatoms, single-atom spin echo requires bringing $\sim N-1$ atoms in a Rydberg state which is prevented by dipole blockade. Finally, unlike discrete or continuous spin echo, driving protection crucially relies on non-Markovianity.

\begin{figure}[tb]
\centering
\includegraphics[width=85mm]{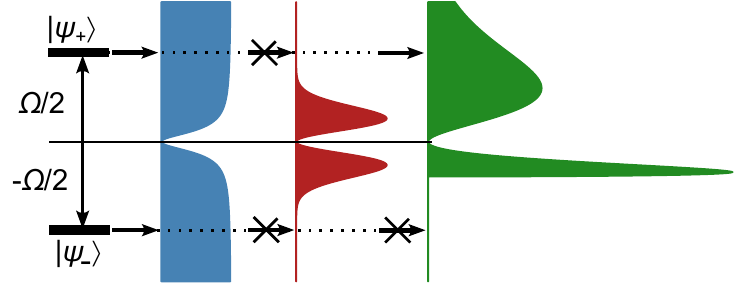}
\caption{ \textbf{Explanation of the driving protection mechanism.} The eigenstates \ket{\psi_\pm} of the driven qubit, shifted by $\pm\Omega/2$, remain resonantly coupled to a ``continuum'' $\{\ket{x}\}$ for a Lorentzian spectrum (blue, left). This coupling becomes off-resonant for a Gaussian spectrum (red, center), which decreases the decay rate of the Rabi oscillations. For a light-shift spectrum, \ket{\psi_+} decays faster than \ket{\psi_-}. }
\label{fig:ExplainDriveProtect}
\end{figure}

The mechanism behind driving protection can be sketched in the following way. In the rotating frame, the eigenstates of the driven collective qubit are $\ket{\psi_\pm}=(\ket{G}\pm\ket{E_0})/\sqrt{2}$, with eigenenergies $\pm\Omega/2$. The dephasing $\sqrt{2}\ox$ couples them to the continuum of states $\{\ket{x}\}$ via $g(x)=\bra{x}\sqrt{2}\ox\ket{\psi_\pm}= \pm x\,\psi_{0}(x+H_0/\sqrt{2})$, the offset $H_0$ accounting for the resonant driving. For a Lorentzian spectrum, $g_\mathrm{M}(x)\propto x/(1+i2x)$ is spectrally broad [Fig.~\ref{fig:ExplainDriveProtect}]: for any $\Omega$,  the states $\ket{\psi_\pm}$ are resonantly coupled to the continuum, which results in irreversible decay. In contrast, for a Gaussian spectrum, $g_\mathrm{G}(x)\propto x\,\ee^{-x^2/2}$ is spectrally narrow and localized around $0$. Strong driving makes the dephasing-induced coupling off-resonant and thus inhibits the decay of the qubit. For the light-shift spectrum, $g_\mathrm{L}(x)\propto x\,\psi_{\mathrm{L},0}(x+3/2)$ is asymmetric, and \ket{\psi_-} is more strongly protected than \ket{\psi_+}.

\section{Experimental methods}
\label{sec:Expmeth}

\begin{figure}[tb]
\centering
\includegraphics[width=85mm]{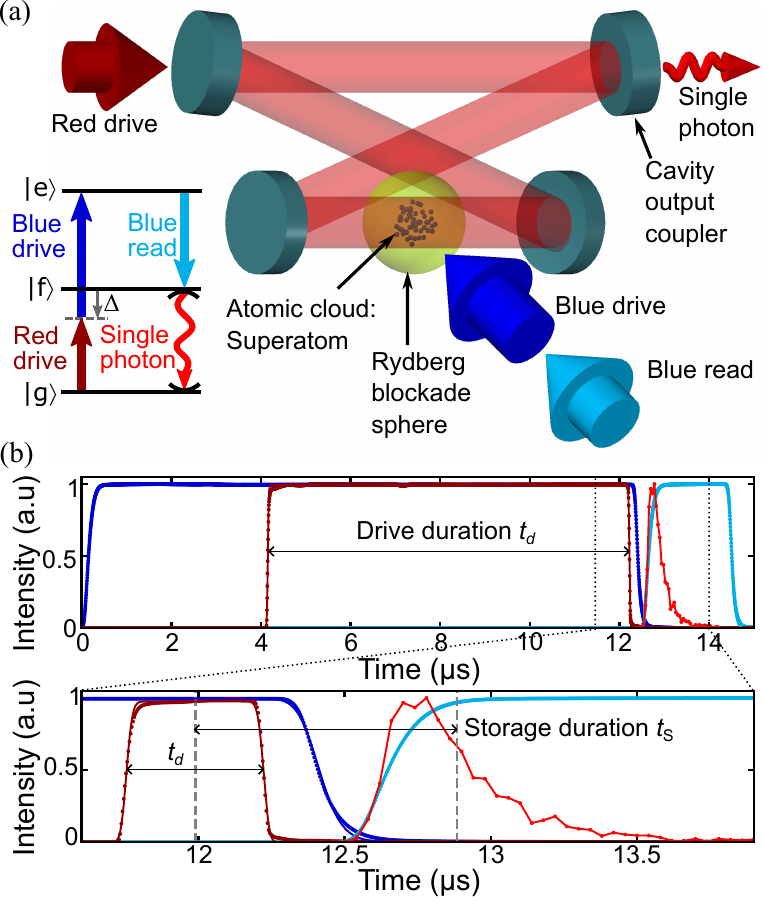}
\caption{ \textbf{Experimental protocol.} (a) Schematic of the experimental setup featuring a Rydberg superatom in an 4-mirror optical cavity, with the laser pulses and the transition scheme used in the experiment, described in Section \ref{sec:Expmeth}. (b) Measured pulse shapes (dots) and best-fit functions (lines) used in the simulations. The lower panel is a zoom on the times between the dotted vertical lines in the upper panel. The blue and dark red curves represent the blue and red drive pulses, respectively. In the lower panel, we represent the short red drive pulse used to prepare the cloud in the singly-excited symmetric Dicke state \ket{E_0} with a fast $\pi$ pulse. The cyan curve represents the blue read beam mapping \ket{E_0} to a photon, while the red line indicates the photon's temporal intensity profile seen by the single-photon detector. The gray dashed lines represent the average times for the $\pi$ pulse and the detected photon.} 
\label{fig:ExpPro}
\end{figure}

To test our predictions experimentally, we use a Rydberg superatom: a small cold atomic ensemble where interactions between Rydberg-state atoms block the number of excitations to one at most, thus defining a collectively-encoded qubit \cite{Pedersen2008,Kumlin2023,Shao2024}. As shown in Fig.~\ref{fig:ExpPro}, the superatom is coupled to a running-wave optical cavity enabling an efficient mapping of its excited state \ket{E_0} to a single-photon Fock state \ket{1}. Therefore, by driving the superatom to its excited state, waiting for a storage duration $t_{\mathrm{s}}$, mapping \ket{E_0} to a photon, and measuring the probability to detect the latter with a photon counter, we obtain a signal proportional to $|s(t_\mathrm{s})|^2$.

The superatom is produced in $\unit{0.1}{\second}$ through the sequential use of two-dimensional (2D) magneto-optical trapping (MOT), 3D MOT, optical molasses cooling, transporting in a one-dimensional optical lattice, and degenerate Raman sideband cooling, as described in detail in Refs.~\cite{Vaneecloo2022, Magro2023}. It is made of $N \simeq 500$ atoms with a Gaussian spatial distribution, loaded in a crossed dipole trap. In the last initialization stage, the atoms are optically pumped in the state $\ket{g} : 5S_{1/2}, F\!\!=\!\!1, m_F\!\!=\!\!1$. By adjusting the depth of the trapping potential during either the preparation or the experimental run, we can control the light-shifts in our system, as well as increase the compression and modify the temperature of the atomic cloud. This allows us to prepare inhomogeneous broadening spectra with similar characteristic widths but with different shapes. To obtain a purely Gaussian Doppler broadening, we switch off the trap during the experimental protocol and increase the confinement potential during the preparation, creating a cloud with a root-mean-square radius  $\sigma_{r} =\unit{3.7(2)}{\micro\meter}$ and a  temperature $T =\unit{6.2(3)}{\micro\kelvin}$. To probe a combination of light-shift and Doppler broadening mechanisms, we do not compress the cloud, which makes it colder ($\sigma_{r} =\unit{4.4(2)}{\micro\meter}$ and $T = \unit{2.5(1)}{\micro\kelvin}$), but apply an average light-shift of $\unit{4.8(6)  10^2}{\kilo\hertz}$ during the experimental protocol. We adjust the temperature and the light shift to achieve similar coherence times for both configurations. Then, we repeat $10$ times the excitation and measurement protocol described below. To remove any residual Rydberg-excited atoms, we apply an electric field pulse using in-vacuum electrodes before each repetition.

In the first ``drive'' step of our protocol, we use a ``red drive'' laser at \unit{795}{\nano\meter} and ``blue drive'' laser at \unit{475}{\nano\meter} to excite a resonant two-photon transition to the Rydberg state $ \ket{e} : 109S, J\!\!=\!\!1/2, m_J\!\!=\!\!1/2 $, with a detuning $\Delta = 2\pi \times \unit{-500}{\mega\hertz}$ from the intermediate state $ \ket{f} : 5P_{1/2}, F\!\!=\!\!2, m_F\!\!=\!\!2$ [see Fig.~\ref{fig:ExpPro}]. 
The $\unit{\sim 20}{\micro\meter}$ Rydberg blockade radius~\cite{Tong2004} largely exceeds the $\unit{4.4}{\micro\meter}$ Gaussian radius of the cloud, strongly inhibiting multiple Rydberg excitations. Thus, the cloud behaves as an effective two-level system with a ground superatom state $\ket{G}$ and a collective singly-excited Rydberg state $\ket{E_0}$. We adjust the collective two-photon Rabi frequency $\Omega = \sqrt{N} \Omega_{\mathrm{r}}\Omega_{\mathrm{b}} / (2 \Delta)$ via the Rabi frequency $\Omega_\mathrm{r}$ of the ``red drive'' beam, keeping constant the Rabi frequency  $\Omega_{\mathrm{b}} = \unit{2\pi\times 7.8(4)}{\mega\hertz}$ of the ``blue drive''. 
The drive duration $t_{\mathrm{d}}$ is set by the ``red drive'' pulse duration, encompassed by the longer ``blue drive'' pulse, as shown in Fig.~\ref{fig:ExpPro}(b). 

During the subsequent ``read'' step, we quasi-adiabatically turn on a ``blue read'' laser resonant with the \ket{e}-\ket{f} transition at \unit{475}{\nano\meter}. As the collective coupling strength $g = 2\pi \times \unit{8.5(3)}{\mega\hertz}$ between the \ket{f}-\ket{g} transition and the resonant cavity mode exceeds the decay rates of both the cavity field ($\unit{2\pi \times 2.9}{\mega\hertz}$) and the atomic dipole ($\unit{2 \pi\times 3}{\mega\hertz}$), this efficiently transfers the excitation from $\ket{E_0}$ to an intra-cavity photon \cite{Gorshkov2007}. This photon then leaks out of the cavity mostly through the output-coupler mirror and is directed to a single-photon detector. 

To measure the free inhomogeneous dephasing of \ket{E_0}, we prepare it by a fast $\pi$ pulse with a strong drive $\Omega/(2\pi) = \unit{1.0(1)}{\mega\hertz}$. Then, we map $\ket{E_0}$ to a photon after a variable storage duration $t_{\mathrm{s}}$, defined as the difference between the average  intensity-weighted times of the $\pi$ pulse and of the extracted photon. To investigate the drive protection, we vary the duration $t_{\mathrm{d}}$ of the drive pulse between $0.1$ and $\unit{11.9}{\micro\second}$ and map $\ket{E_0}$ to a photon directly afterwards, revealing point by point the Rabi oscillations.

\begin{figure}[tb]
\centering
\includegraphics[width=85mm]{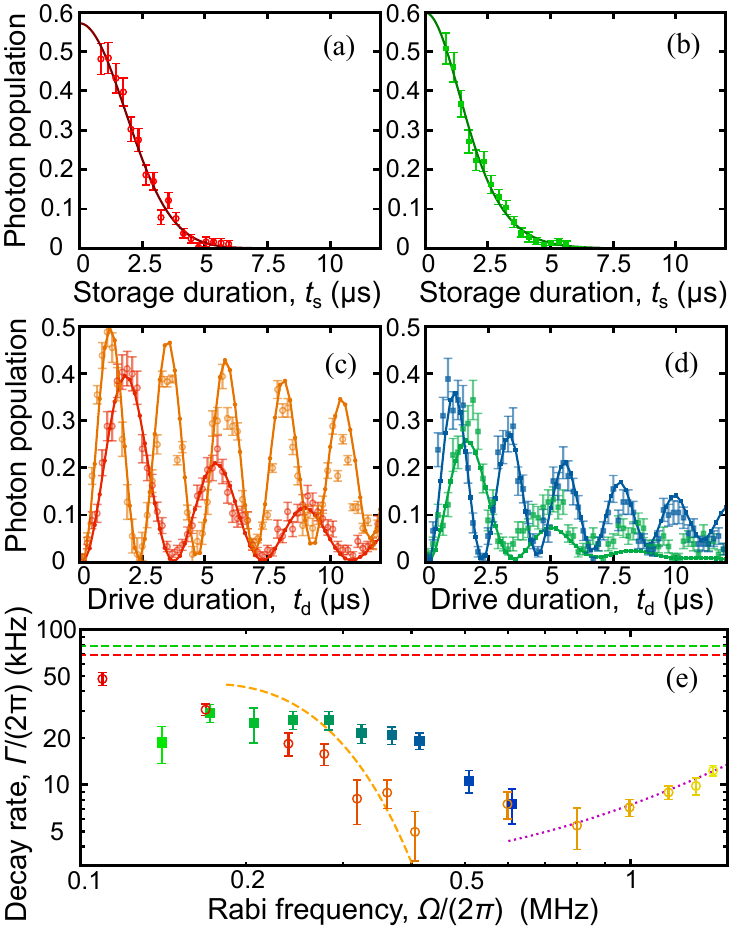}
\caption{ \textbf{Experimental observation of driving protection in a Rydberg superatom.} The detected photon population is proportional to the population $|s(t)|^2$ of the qubit's excited state \ket{E_0}. We realize two dephasing spectra: purely Gaussian, due to thermal Doppler broadening (left (a)(c) and red to yellow colored circles) and a convolution of a light-shift in a slightly anharmonic trap with a Gaussian due to a residual Doppler broadening (right (b)(d) and green to blue colored squares). Error bars represent the standard error. (a) and (b): Free evolution of the excited state \ket{E_0}, fitted with the corresponding rescaled $|s(t)|^2$ functions, yielding mean lifetimes of $\unit{2.33(7)}{\micro\second}$ and $\unit{2.03(9)}{\micro\second}$, respectively. (c) and (d): Evolution with Rabi frequencies $\Omega = \unit{2 \pi \times (239, 406, 244, 414)}{\kilo\hertz}$ for (red, orange, green, blue), respectively. Points connected by lines show the results of corresponding numerical simulations using our dephasing model, parametrized with independently measured values. (e): Fitted exponential decay rate $\Gamma$ of the observed oscillation amplitude [fit function of Eq.~\ref{eq:fluctRabiFreq} in Appendix~\ref{app:simDrive}] as a function of the Rabi frequency $\Omega/(2\pi)$. The horizontal dashed red and green lines show inverses of the free-decay lifetimes given by fits in (a) and (b), respectively. The dashed orange line is the expression of $\gamma_{\mathrm{G},\Omega}$ given by Eq.~\ref{eq:SGOmegaAppr}. The dotted purple line is a best-fitted $\Gamma_{\Omega} \, \Omega^2 + \Gamma_0$ function for points where $\Omega > \unit{2 \pi \times 0.7}{\mega\hertz}$. }
\label{fig:exp}
\end{figure}

\section{Experimental results}
\label{sec:expRes}

The results of the experiment are presented in Fig.~\ref{fig:exp}. Panels (a) and (b) show the decay of $\ket{E_0}$ without driving, fitted with the solutions of Eq.\ref{EqDecaySvsX}. The Gaussian decay in panel (a), with a mean lifetime $ \sqrt{\pi}/(2 \omega_{\mathrm{G}}) = \unit{2.33(7)}{\micro\second}$, results from pure Doppler thermal broadening. Panel (b) shows a decay with a similar mean lifetime of $\unit{2.03(9)}{\micro\second}$, but with a different shape, arising from a convolution between a Gaussian, due to residual thermal motion, and a light-shift spectrum slightly modified by the anharmonicity of our optical dipole trap [details in Appendix~\ref{app:simulations}]. 

Panels (c) and (d) correspond to the same configurations in presence of Rabi flopping. They confirm the predicted driving protection: the visibility of Rabi oscillations increases with their frequency and reaches coherence times strongly exceeding the free decay lifetimes measured above.  The observed dynamics are reproduced by numerical simulations on a desktop computer, using a truncated basis of Dicke states [details in Appendix~\ref{app:simulations}]. By fitting an exponential envelope $\ee^{- \Gamma t_{\mathrm{d}}}$ on the measured oscillations, with a fit function accounting for shot-to-shot fluctuations of the number of atoms in the prepared superatom [details in Appendix~\ref{app:simDrive}], we obtain the decay rate $\Gamma$ presented in panel (e). It decreases rapidly with the Rabi frequency, down to a factor $14(5)$ below the characteristic free decay rate of \ket{E_0}.  As expected from Sec.~\ref{sec:DecoFree}, at intermediate values $ 5 < \Omega /\omega_{\mathrm{G}} < 10 $ (i.e $\unit{0.3}{\mega\hertz} < \Omega/(2\pi) < \unit{0.6}{\mega\hertz}$), the driving protection is less efficient in the light-shifted case than in the Gaussian one. For the latter, the data points are close to the analytic result of Eq.~\ref{eq:SGOmegaAppr}, valid in the strong driving limit and plotted for $\Omega > 3 \omega_{\mathrm{G}}$. At very strong driving, $\Gamma$ is ultimately limited by experimental effects compatible with a form $\Gamma_{\Omega} \, \Omega^2 + \Gamma_0$ \cite{Finkelstein2021}. $\Gamma_{\Omega}$ accounts for decoherence mechanisms proportional to the power of the driving beam, including spontaneous emission of the intermediate level of the two-photon transition between \ket{G} and \ket{E_0} and imperfect Rydberg blockade; while $\Gamma_0$ stems from defects such as the finite lifetime of the Rydberg state, imperfect optical pumping, impure laser polarizations, or residual electric fields.

\section*{Conclusion}

We unveiled, theoretically and experimentally, a dynamical decoupling effect for collective qubits, which enhances coherence times by a factor $14$ in our experimental system.   
The demonstration of this driving protection effect was enabled by the simplicity of the underlying theoretical framework. Among existing descriptions of non-Markovian quantum dynamics \cite{Breuer2016,DeVega2017}, it has some resemblance with the formalism of pseudomodes \cite{Mazzola2009}. Applicable to a large class of physical systems, it gives access to an easy implementation, to efficient numerics, and to intuitive physical interpretations exemplified here.

It is worth comparing this protection mechanism with other ways of counteracting inhomogeneous dephasing, all of which come at some practical cost. Firstly, one can act on the dephasing's causes. For the Doppler and the light-shift broadenings, it means cooling the atoms further, which significantly degrades experimental duty cycles, and switching off the trapping light, which eventually leads to atomic loss. Alternatively, the system can be made less sensitive to dephasing, by using Doppler-free excitation or state-insensitive dipole traps. For Rydberg superatoms, such traps operate close to atomic transitions \cite{Lampen2018}, while Doppler-free excitation requires three laser beams with specific angles  \cite{Ryabtsev2011}, putting significant constrains on optical access. Other mitigation schemes for Doppler broadening in Rydberg gases involve a continuous off-resonant coupling \cite{Finkelstein2021} or a transient transfer \cite{Jiao2025} to an auxiliary state. Compared to these schemes, the strengths and weaknesses of driving protection are similar to those of spin echo, applicable to collectively-encoded qubits \cite{Rui2015} but, as mentioned earlier, incompatible with strong Rydberg blockade. It leads to unwanted phase accumulations, sensitive to fluctuations of the Rabi frequency, but it is not restricted to specific ranges of atomic levels and it requires no resources other than those already available to manipulate the qubit.

Besides its potential applications in quantum information processing with collective qubits, the driving protection process is interesting in its own right. Compared to cavity protection, where the collective coupling between the cavity and the emitters is generally fixed \cite{Baghdad2023}, the temporal control of the collective Rabi frequency gives access to dynamics yet to be explored, such as the active stabilization of the eigenstates of the coupled system.

\section*{Funding}
This work was funded by the ERC Starting Grant 677470 SEAQUEL, the Plan France 2030 through the project ANR-23-PETQ-0013, the Ile-de-France DIM network SIRTEQ through the project CRIMP, and the CIFAR Azrieli Global Scholars program.

\section*{Acknowledgements}
We thank J. Vaneecloo for his help during early experimental developments and characterizations.

\section*{Disclosures}
The authors declare no conflicts of interest.

\section*{Data availability}
Data underlying the results presented in this paper are available upon reasonable request from the authors.


%

\pagebreak
\onecolumngrid
\appendix
\counterwithin{figure}{section}

\section{Lamb-like shift and decay rate}
\label{app:decay}

In the model described in the
 manuscript, we identify a ladder of Dicke states $\{\ket{E_k}\}$ sequentially coupled by the inhomogeneous-dephasing Hamiltonian $\oH=\sqrt{2}\ox$, and calculate the diagonal terms $H_k = \langle E_k | \oH | E_k \rangle$ and the coupling terms $V_k = \langle E_{k-1} | \oH | E_k \rangle$. We then restrict the system's Hilbert space to the finite subspace $\{\ket{E_{k\leq n}}\}$ formed by the lowest $n+1$ states, and treat the orthogonal subspace spanned by $\{\ket{E_{k> n}}\}$ as a broad continuum. Here we calculate the amplitude decay rate $\gamma_n$ and the Lamb-like energy shift $\delta_n$ of \ket{E_n} induced by the coupling to this continuum. For this, we define the 
projectors $\oP=\sum_{k=0}^{n}\ketbra{E_k}{E_k}$ and $\oQ=1-\oP$ into the two subspaces of interest, and derive an effective non-Hermitian Hamiltonian $\oH_\mathrm{e}$ from the resolvent $\oG(z)=(z-\oH)^{-1}$ which obeys \cite{Cohen1998.ch3} 
\begin{align}
\label{EqHeff}
\oP\oG(z)\oP &= \frac{1}{z\oP-\oH_\mathrm{e}(z)},&
\oH_\mathrm{e}(z) &= \oP\oH\oP+\oP\oH\oQ\frac{1}{z\oQ-\oQ\oH\oQ}\oQ\oH\oP.
\end{align}
To obtain a parameter-free $\oH_\mathrm{e}$, the energy $z$ is fixed to a physically-relevant value. As \ket{E_0} is the only experimentally-accessible state, we set $z$ to $H_0$. Since $\oH$ is tridiagonal in this basis, 
\begin{align}
\oH_\mathrm{e}(z) &= \oP\oH\oP+\Delta_n\ketbra{E_n}{E_n},&
\Delta_n &= \delta_n-i\gamma_n= |V_{n+1}|^2\bra{E_{n+1}}\oG_{n+1}(z)\ket{E_{n+1}}
\end{align}
where $\oG_{n+1}(z)=(z\oQ-\oQ\oH\oQ)^{-1}$ is the resolvent of $\oQ\oH\oQ$. $\Delta_n$ can be calculated inductively by replacing $G$ with $\oG_n$ and $\oP$ with $\oP_n=\ketbra{E_{n}}{E_{n}}$ in Eq.~\ref{EqHeff}:
\begin{align}
\label{eq:appDeltaRec}
\oP_{n}\oG_{n}(z)\oP_{n} =& \oP_{n}\frac{\Delta_{n-1}}{|V_n|^2}=\frac{\oP_n}{z-H_n-\Delta_n}&
\Rightarrow \Delta_n =& z-H_n-\frac{|V_n|^2}{\Delta_{n-1}}.
\end{align}
The initial value $\Delta_0$ is given by the limit $\Imag(z)\rightarrow 0^+$ of
\begin{align}
\label{eq:G0}
G_0(z)
= \frac{1}{z-H_0-\Delta_0}
= \bra{E_0}\oG(z)\ket{E_0}
= \int \frac{\rho(x)\dd x}{z-\sqrt{2} x}
= -i \int_0^{+\infty}\ee^{izt}s(t)\dd t,
\end{align}
the noise spectral density $\rho(x)$ and the amplitude $s(t)=\langle E_0|\ee^{-i\oH t}|E_0\rangle=\int\dd x \rho(x)\ee^{-i\sqrt{2}xt}$ being defined in the main text.

\medskip
We can now apply this procedure to the three specific cases discussed in the manuscript.
\begin{itemize}
\item
For the Markovian case $\rho_\mathrm{M}$ where $H_{\mathrm{M},0}=0$,
\begin{align}
\label{EqGM0}
s_\mathrm{M}(t)
=&
\int_{-\infty}^{+\infty}\dd x \frac{2\ee^{-i\sqrt{2}xt}}{\pi(1+4x^2)}
= \frac{1}{2i\pi} \int_{-\infty}^{+\infty}\dd x \ee^{-ixt}\left(\frac{1}{x-i/\sqrt{2}}-\frac{1}{x+i/\sqrt{2}}\right)=\ee^{-t/\sqrt{2}} \\
\Rightarrow G_{\mathrm{M},0}(z)
=& -i\int_0^{+\infty}\dd t \ee^{-t/\sqrt{2}+itz}
=\frac{1}{z+i/\sqrt{2}}.
\end{align} 
Self-consistently, this gives $\delta_{\mathrm{M},0}=0$ and $\gamma_{\mathrm{M},0}=1/\sqrt{2}$ independently on $z$.
\item
For the Gaussian spectrum $\rho_\mathrm{G}$, 
\begin{align}
s_\mathrm{G}(t)
=& \frac{1}{\sqrt{\pi}}\int_{-\infty}^{+\infty}\dd x \ee^{-x^2-i\sqrt{2}xt}
=\ee^{-t^2/2},\\
\label{eq:G0G}
\Rightarrow G_{\mathrm{G},0}(z)
=& - i\int_{0}^{+\infty}\dd t\; \ee^{-t^2/2 + itz}
= - i\sqrt{2}\ee^{-z^2/2}\int_{-iz/\sqrt{2}}^{+\infty}\dd t\; \ee^{-t^2}
= - i\sqrt{\frac{\pi}{2}} w\left(\frac{z}{\sqrt{2}}\right),
\end{align}
where $w(z)=\ee^{-z^2}\erfc(-iz)$ is the Faddeeva function. With $z=H_{\mathrm{G},0}=0$, we get $\Delta_{\mathrm{G},0}=-1/G_{\mathrm{G},0}(0)=-i\sqrt{2/\pi}$ and, from Eq.~\ref{eq:appDeltaRec},
\begin{align}
\label{EqDeltaEffG}
\Delta_{\mathrm{G},n} = \frac{n}{\Delta_{\mathrm{G},n-1}} = \frac{n}{n-1}\Delta_{\mathrm{G},n-2}=...
&& \Rightarrow &&
\delta_{\mathrm{G},n}
= 0,
&&
\gamma_{\mathrm{G},n} 
= \frac{(n)!!}{(n-1)!!}\left(\frac{2}{\pi}\right)^{(-1)^n/2}.
\end{align}
\item
For the light-shift spectrum $\rho_\mathrm{L}$, assuming $\Imag(z)\rightarrow 0^+$,
\begin{align}
\nonumber
G_{\mathrm{L},0}(z)
=& \frac{2}{\sqrt{\pi}}\int_{0}^{+\infty}\dd x \frac{\sqrt{x}\ee^{-x}}{z-\sqrt{2} x}
= -\sqrt{\frac{2}{\pi}}\int_{-\infty}^{+\infty}\dd u\frac{u^2\ee^{-u^2}}{u^2-z/\sqrt{2}}\\
\nonumber
=& -\sqrt{2} - \sqrt{\frac{2}{\pi}} \int_{-\infty}^{+\infty}\dd u\ee^{-u^2}\frac{1}{2}\sqrt{\frac{z}{\sqrt{2}}}\left(\frac{1}{u-\sqrt{z/\sqrt{2}}}-\frac{1}{u+\sqrt{z/\sqrt{2}}}\right)\\
\nonumber
=&  -\sqrt{2}- i\sqrt{z\sqrt{2}} \frac{1}{\sqrt{\pi}} \int_{-\infty}^{+\infty}\dd u\int_{0}^{+\infty}\dd t\ee^{-u^2-it(u-\sqrt{z/\sqrt{2}})}
=  -\sqrt{2}- i\sqrt{z\sqrt{2}} \int_{0}^{+\infty}\dd t\ee^{-t^2/4+it\sqrt{z/\sqrt{2}}}\\
\label{eq:G0L}
=& -\sqrt{2}-i\sqrt{\pi}\sqrt{z\sqrt{2}}w\left(\sqrt{\frac{z}{\sqrt{2}}}\right).
\end{align}
With $z=H_{\mathrm{L},0}=3/\sqrt{2}$, $\Delta_{\mathrm{L},0}=1/(\sqrt{2}+i\sqrt{3\pi}w(\sqrt{3/2}))$. Eq.~\ref{eq:appDeltaRec} becomes $\Delta_n=-2^{3/2}n-n(2n+1)/\Delta_{n-1}$ and gives $\delta_n$ and $\gamma_n$ by induction.
\end{itemize}

\section{Driving protection mechanism}
\label{app:protection}

Let us consider here that the collective qubit is resonantly driven between \ket{G} and \ket{E_0} with a constant Rabi frequency $\Omega$, assumed to be real. In the rotating frame, writing $\os_{G0}=\ket{G}\bra{E_0}$,  the Hamiltonian becomes
\begin{align}
\oH_\Omega=&\oH_\mathrm{r}+\oD
& \oH_\mathrm{r}
=&\sqrt{2}\ox-H_0(1-\os_{GG}),
& \oD=\frac{\Omega}{2}(\os_{G0}+\os_{0G})
\end{align}
where the dephasing term $\oH_\mathrm{r}$ is offset by $H_0$ to make the driving $\oD$ resonant. 
We define the resolvents $\oG_\mathrm{r}$ and $\oG_\Omega$ of $\oH_\mathrm{r}$ and $\oH_\Omega$, 
and use the general relation $\hat{B}^{-1}=\hat{A}^{-1}+\hat{A}^{-1}(\hat{A}-\hat{B})\hat{B}^{-1}$ to rewrite \cite{Cohen1998.ch3}
\begin{align}
\label{EqSeriesG}
\oG_\Omega(z) =& \oG_\mathrm{r}(z)+\oG_\mathrm{r}(z)\oD\oG_\Omega(z)
= \oG_\mathrm{r}(z)+\oG_\mathrm{r}(z)\oD\oG'(z) + \oG_\mathrm{r}(z)\oD\oG'(z)\oD\oG_\Omega(z) = ...
= \sum_{n=0}^{+\infty} \left(\oG_\mathrm{r}(z)\oD\right)^n\oG_\mathrm{r}(z).
\end{align}
As $\ox\ket{G}=0$, we have
\begin{align}
\oG_\mathrm{r}(z)\oD
=&\frac{\Omega}{2}\left(\frac{1}{z}\os_{G0}+\frac{1}{z-\sqrt{2}\ox+H_0(1-\os_{GG})}\os_{0G}\right),\\
\left(\oG_\mathrm{r}(z)\oD\right)^2
=&\left(\frac{\Omega}{2}\right)^2\frac{1}{z}\left(G_0(z+H_0)\os_{GG}+\frac{1}{z-\sqrt{2}\ox+H_0(1-\os_{GG})}\os_{00}\right),\\
\left(\oG_\mathrm{r}(z)\oD\right)^3
=&\left(\frac{\Omega}{2}\right)^3\frac{G_0(z+H_0)}{z}\left(\frac{1}{z}\os_{G0}+\frac{1}{z-\sqrt{2}\ox+H_0(1-\os_{GG})}\os_{0G}\right)
= \frac{\Omega^2 G_0(z+H_0)}{4z} \oG_0(z)\oD\\
\Rightarrow 
\left(\oG_\mathrm{r}(z)\oD\right)^{2n+k}
=&\left(\frac{\Omega^2 G_0(z+H_0)}{4z} \right)^n \left(\oG_\mathrm{r}(z)\oD\right)^{k},\\
\label{EqGvsG0}
\Rightarrow 
\oG_\Omega(z)
=&\oG_\mathrm{r}(z)+\frac{\oG_\mathrm{r}(z)\oD\oG_\mathrm{r}(z)+(\oG'(z)\oD)^2\oG_\mathrm{r}(z)}{1-\Omega^2 G_0(z+H_0)/(4z)}
\end{align}
where $G_0$ is given by Eq.~\ref{eq:G0}. The forward propagator  \cite{Cohen1998.ch3}
\begin{equation}
\label{eq:UOmega}
\oU_\Omega(t>0)
= \ee^{-i\oH_\Omega t}
= \lim_{\eta\rightarrow 0^+}\int_{-\infty}^{+\infty}\frac{i\dd z}{2\pi}\oG_\Omega(z+i\eta)\ee^{-itz},
\end{equation}
gives access to the dynamics of the qubit through the restriction of $\oG_\Omega$ to the subspace $\{\ket{G},\ket{E_0}\}$,
\begin{equation}
\label{eq:GSeriesProj}
\oG_{\Omega,0}(z) = \frac{\os_{GG}+G_0(z+H_0)(z\,\os_{00}+\oD)}{z- \frac{\Omega^2}{4}G_0(z+H_0)}.
\end{equation}
In particular, the amplitude of the qubit's excited state now evolves as 
\begin{equation}
\label{eq:appSOmega}
s_\Omega(t)
= \bra{E_0}\oU_\Omega(t)\ket{E_0}
= \int_{-\infty}^{+\infty}\frac{i\dd z}{2\pi}\frac{zG_0(z+H_0)\ee^{-itz}}{z-\frac{\Omega^2}{4}G_0(z+H_0)}.
\end{equation} 
For $\Omega\rightarrow 0$ we recover the undriven case $s(t)$ discussed in the theoretical section of the main manuscript, up to a phase factor $\ee^{iH_0t}$ corresponding to the resonance condition. In the following we focus on the opposite strong-driving regime where $\Omega\gg 1$.

\subsection{Markovian process}

For a Lorentzian spectrum $\rho_\mathrm{M}$, the simple form of $G_{\mathrm{M},0}$ in Eq.\ref{EqGM0} gives
\begin{align}
s_{\mathrm{M},\Omega}(t)
=&
\int_{-\infty}^{+\infty}\frac{i\dd z}{2\pi}\frac{z\ee^{-itz}}{z(z+i\gamma_\mathrm{M})-\frac{\Omega^2}{4}}.
\end{align}
where $\gamma_\mathrm{M}=1/\sqrt{2}$ is the driving-less decay rate derived in the theoretical section of the manuscript. Writing $ \Omega_\mathrm{M} = \sqrt{\Omega^2-\gamma_\mathrm{M}^2}$,
\begin{multline}
s_{\mathrm{M},\Omega}(t)
=
\int_{-\infty}^{+\infty}\frac{i\dd z}{2\pi}\frac{\ee^{-itz}}{2}\left(\frac{1-i\gamma_\mathrm{M}/\Omega_\mathrm{M}}{z-\Omega_\mathrm{M}/2+i\gamma_\mathrm{M}/2}+ \frac{1+i\gamma_\mathrm{M}/\Omega_\mathrm{M}}{z+\Omega_\mathrm{M}/2+i\gamma_\mathrm{M}/2} \right)\\
= \ee^{-\gamma_\mathrm{M}t/2}\left(\cos\left(\frac{\Omega_\mathrm{M}t}{2}\right) - \frac{\gamma_\mathrm{M}}{\Omega_\mathrm{M}} \sin\left(\frac{\Omega_\mathrm{M}t}{2}\right) \right).
\end{multline}
When $\Omega>\gamma_\mathrm{M}$, the decay rate $\gamma_\mathrm{M}/2$, two times smaller than without driving, becomes independent on the driving strength.

We can verify these results using standard Bloch equations. For consistency with the model above, let us assume that the qubit oscillates between the states \ket{g} and \ket{e} and decays into a different state \ket{d}. The master equation
\begin{align}
\frac{\dd}{\dd t}\orho =&-i\frac{\Omega}{2}\left[\os_{ge}+\os_{eg},\orho\right]
+\gamma\left(2\os_{de}\orho\os_{ed}-\os_{ee}\orho-\orho\os_{ee}\right)
\end{align}
leads to 
\begin{equation}
\frac{\dd}{\dd t} \left(\begin{array}{c} \orho_{ee} \\ \Imag\orho_{ge} \\ \orho_{dd} \end{array} \right)
=
-\left(\begin{array}{ccc}
2\gamma & -\Omega & 0 \\
\Omega & \gamma & \Omega/2\\
-2\gamma & 0 & 0
\end{array} \right)
\left(\begin{array}{c} \orho_{ee} \\ \Imag\orho_{ge} \\ \orho_{dd} \end{array} \right)
+ \left(\begin{array}{c} 0 \\ \Omega/2 \\ 0 \end{array} \right).
\end{equation}
For $\Omega>\gamma$, the real parts of the eigenvalues $\{\gamma,\gamma\pm i\sqrt{\Omega^2-\gamma^2}\}$ of the matrix are all equal to $\gamma$, independently on $\Omega$. Setting $\gamma=\gamma_\mathrm{M}$ allows one to recover the results above.

\subsection{Gaussian process}

For a Gaussian spectrum, $H_{\mathrm{G},0}=0$ and  $G_{\mathrm{G},0}(z)$ is given by Eq.~\ref{eq:G0G}, which can be rewritten as
\begin{align}
G_{\mathrm{G},0}(z) =& \sqrt{2}D\left(\frac{z}{\sqrt{2}}\right)-i\sqrt{\frac{\pi}{2}}\ee^{-z^2/2},&
D(z) =& \ee^{-z^2}\int_0^z\ee^{t^2}\dd t,
\end{align}
where $D(z)$ is the  Dawson integral. The even function $D(z)/z$ monotonically decreases from $1$ to $0$  on $\mathbb{R}_+$, which means that for $\Omega\gg 1$ the real part of the denominator in the integrand of Eq.~\ref{eq:appSOmega} vanishes in $z_\pm = \pm\Omega_\mathrm{G}/2$ with $\Omega_\mathrm{G} \gg 1$, where its imaginary part becomes very small. We can recursively user the asymptotic expansion 
\begin{equation}
\Real(G_{\mathrm{G},0}(z))=\sum_{k=0}^N\frac{(2k-1)!!}{z^{2k+1}}+O\left(\frac{1}{z^{2N+3}}\right),
\end{equation}
to solve $\Omega_\mathrm{G}/2=(\Omega^2/4)\Real(G_{\mathrm{G},0}(\Omega_\mathrm{G}/2))$
and find $\Omega_\mathrm{G}$ to an arbitrary order in $\Omega$. To the lowest non-trivial order,
 $\Omega_\mathrm{G} = \Omega+2/\Omega+O(\Omega^{-3})$. Near $z_\pm$, to the first order in $z-z_\pm$ and in $1/\Omega$,
\begin{equation}
z-\frac{\Omega^2}{4}\Real((G_{\mathrm{G},0}(z)) 
\approx (z-z_\pm)\left(1-\frac{\Omega^2}{4} \Real((G_{\mathrm{G},0}'(z_\pm)) \right)
\approx  (z-z_\pm)\left(1+\frac{\Omega^2}{4} \frac{1}{z_\pm^2} \right)
\approx 2(z-z_\pm).
\end{equation} 
As the imaginary part of the integrand's denominator is suppressed exponentially in $\Omega^2$ by $\Imag(G_{\mathrm{G},0}(z_\pm))=-\sqrt{\pi/(2e)}\ee^{-\Omega^2/8}(1+O(\Omega^{-2}))$, we find
\begin{align}
\frac{zG_{\mathrm{G},0}(z)}{z-\frac{\Omega^2}{4}G_{\mathrm{G},0}(z)}
\approx 
\frac{1}{2}\left(\frac{1}{z-\Omega_\mathrm{G}/2+i\gamma_{\mathrm{G},\Omega}} + \frac{1}{z+\Omega_\mathrm{G}/2+i\gamma_{\mathrm{G},\Omega}}\right),&&
\Omega_\mathrm{G}
\approx 
\Omega+\frac{2}{\Omega},&&
\gamma_{\mathrm{G},\Omega} \approx \frac{\Omega^2}{8}\sqrt{\frac{\pi}{2e}}\ee^{-\Omega^2/8}.
\end{align}
Eq.~\ref{eq:appSOmega} then integrates into $s_\mathrm{G}(t)\approx\ee^{-\gamma_{\mathrm{G},\Omega}t}\cos(\Omega_\mathrm{G}t/2)$, showing that strong driving turns the Gaussian decay into an exponential one, and that the rate $\gamma_\mathrm{G}$ of this exponential decay decreases extremely fast when $\Omega$ increases. Even for a moderate $\Omega=10$, $\gamma_\mathrm{G}$ drops below $10^{-4}$, allowing Rabi oscillations to persist for much longer that the decay time of he undriven qubit.

\subsection{Light shift in a harmonic trap}

The case of the harmonic light shift spectrum $\rho_\mathrm{L}$ resembles the Gaussian one. Here $H_{\mathrm{L},0}=3/\sqrt{2}$ and the asymptotic expansion of the real part of Eq.~\ref{eq:G0L} gives
\begin{align}
\Real (G_{\mathrm{L},0}(z))
=&
\frac{1}{z} \sum_{k=0}^{N}\frac{(2k+1)!!}{(\sqrt{2}z)^k}+O\left(\frac{1}{z^{N+2}}\right)
&
\Rightarrow \Real (G_{\mathrm{L},0}(z+H_{\mathrm{L},0}))
=&
\frac{1}{z}+\frac{3}{z^3}+O\left(\frac{1}{z^4}\right).
\end{align}
This shows that real part of the denominator of the integrand in Eq.~\ref{eq:appSOmega} vanishes in two nearly opposite points $z_\pm=\pm(\Omega/2+3/\Omega)+12\sqrt{2}/\Omega^2+O(1/\Omega^3)$. Using $w(iz)=\ee^{z^2}\erfc(z)$ to rewrite
\begin{equation}
G_{\mathrm{L},0}(-z) = -\sqrt{2}+\sqrt{\pi z\sqrt{2}}\ee^{z/\sqrt{2}}\erfc\left(\sqrt{\frac{z}{\sqrt{2}}}\right),
\end{equation}
we see that, for $z>0$, $\Imag(G_{\mathrm{L},0}(z))=-\sqrt[4]{2}\sqrt{\pi z}\ee^{-z/\sqrt{2}}$ but $\Imag(G_{\mathrm{L},0}(-z))=0$. With $G_{\mathrm{L},0}'(z+H_{\mathrm{L},0})=-z^{-2}+O(z^{-4})$, the same approximations as in the Gaussian case give again
\begin{align}
z-\frac{\Omega^2}{4}\Real((G_{\mathrm{L},0}(z+H_{\mathrm{L},0})) 
\approx & (z-z_\pm)\left(1-\frac{\Omega^2}{4} \Real((G_{\mathrm{L},0}'(z_\pm)) \right)
\approx  (z-z_\pm)\left(1+\frac{\Omega^2}{4} \frac{1}{z_\pm^2} \right)
\approx 2(z-z_\pm).
\end{align}
Near $z_+$, the imaginary part of the integrand's denominator now decreases exponentially with $\Omega$, while near $z_-$ it vanishes:
\begin{align}
\frac{\Omega^2}{4}\Imag((G_{\mathrm{L},0}(z_++H_{\mathrm{L},0}))
= 
\frac{\Omega^2}{4}\sqrt{\frac{\pi\Omega}{\sqrt{2}e^3}}\ee^{-\Omega/\sqrt{8}}(1+O(\Omega^{-2})),
&&
\frac{\Omega^2}{4}\Imag((G_{\mathrm{L},0}(z_-+H_{\mathrm{L},0})) = 0
\end{align}
This gives
\begin{align}
\frac{z G_{\mathrm{L},0}(z)}{z-\frac{\Omega^2}{4}G_{\mathrm{L},0}(z)}
\approx 
\frac{1}{2}\left(\frac{1}{z-\Omega_\mathrm{L}/2+i\gamma_{\mathrm{L},\Omega}} + \frac{1}{z+\Omega_\mathrm{L}/2}\right),&&
\Omega_\mathrm{L}
\approx 
\Omega+\frac{6}{\Omega},&&
\gamma_{\mathrm{L},\Omega} \approx \frac{\Omega^2}{8}\sqrt{\frac{\pi\Omega}{\sqrt{2}e^3}}\ee^{-\Omega/\sqrt{8}},
\end{align}
and integrates into 
\begin{equation}
s_\mathrm{L}(t) =\bra{E_0}\oU_{\mathrm{L},\Omega}(t)\ket{E_0} \approx (\ee^{-\gamma_{\mathrm{L},\Omega}t}\ee^{-i\Omega_\mathrm{L}t/2}+\ee^{i\Omega_\mathrm{L}t/2})/2.
\end{equation}

Using Eqs.~\ref{eq:UOmega} and \ref{eq:GSeriesProj}, we can apply the same procedure to the three other elements of the restricted propagator $\oU_{\mathrm{L},\Omega}(t>0)$:
\begin{align}
\bra{G}\oU_{\mathrm{L},\Omega}(t)\ket{G}
=&
\int_{-\infty}^{+\infty}\frac{i\dd z}{2\pi}\frac{\ee^{-itz}}{z-\frac{\Omega^2}{4}G_{\mathrm{L},0}(z+H_{\mathrm{L},0})}
\approx
\frac{\ee^{-\gamma_{\mathrm{L},\Omega}t}\ee^{-i\Omega_\mathrm{L}t/2}+\ee^{i\Omega_\mathrm{L}t/2}}{2}
= \bra{E_0}\oU_{\mathrm{L},\Omega}(t)\ket{E_0},\\
\bra{E_0}\oU_{\mathrm{L},\Omega}(t)\ket{G}
=&
\int_{-\infty}^{+\infty}\frac{i\dd z}{2\pi}\frac{\frac{\Omega}{2} G_{\mathrm{L},0}(z+H_{\mathrm{L},0})\ee^{-itz}}{z-\frac{\Omega^2}{4}G_{\mathrm{L},0}(z+H_{\mathrm{L},0})}
\approx
\frac{\ee^{-\gamma_{\mathrm{L},\Omega}t}\ee^{-i\Omega_\mathrm{L}t/2} -\ee^{i\Omega_\mathrm{L}t/2}}{2}
= \bra{G}\oU_{\mathrm{L},\Omega}(t)\ket{E_0}.
\end{align}
Thus, in the strong driving limit, the propagator restricted to the qubit's subspace can be written as 
\begin{equation}
\oU_{\mathrm{L},\Omega}(t) \approx
 \ee^{-\gamma_{\mathrm{L},\Omega}t}\ee^{-i\Omega_\mathrm{L}t/2}\ketbra{\psi_+}{\psi_+}
+ \ee^{i\Omega_\mathrm{L}t/2}\ketbra{\psi_-}{\psi_-}
\end{equation}
where $\psi_{\pm}=(\ket{G}\pm\ket{E_0})/\sqrt{2}$ are the eigenstates of the Rabi flopping Hamiltonian $\oD$. The ``driving protection'' of $\ket{\psi_+}$ is weaker than for the Gaussian spectrum, as $\gamma_{\mathrm{L},\Omega}$ decreases exponentially with $\Omega$ instead of $\Omega^2$, while $\ket{\psi_-}$ does not decay at all.

\section{Simulations}
\label{app:simulations}

The simulations presented together with our experimental results in the main manuscript use the Python package QuTiP~\cite{Johansson2013} to  integrate the GKSL master equation
\begin{equation}
\frac{\dd \hat{\rho}}{\dd t} = -i\left[\hat{H}_{\mathrm{Drive}} + \hat{H}_{\mathrm{Deph}},\hat{\rho}\right] +\mathcal{L}_{\mathrm{Deph}}[\hat{\rho}]+\mathcal{L}_{\mathrm{Loss}}[\hat{\rho}] ,
\end{equation}
incorporating the following terms.

\subsection{Driving term $\hat{H}_{\mathrm{Drive}}$ and fluctuations of $\Omega$.}

\label{app:simDrive}
In $\hat{H}_{\mathrm{Drive}} = (\Omega/2)(\ket{E_0}\bra{G} +\ket{G}\bra{E_0})$, describing the Rabi flopping of the qubit, the temporal profile of the collective Rabi frequency $\Omega=\sqrt{N}\Omega_\mathrm{r}\Omega_\mathrm{b}/(2\Delta)$ is obtained from ``red drive'' and ``blue drive'' laser pulse shapes measured with photodiodes and plotted in Fig.~5(b) in the main manuscript. The detuning $\Delta$ and the ``blue drive'' Rabi frequency $\Omega_\mathrm{b}$ being fixed (see ``Experimental Methods'' in the manuscript), the proportionality factor between $|\Omega|^2$ and the controlled power of the ``red drive'' beam is extracted from experimental Rabi oscillations in the regime where $\Omega \gg \omega_{\mathrm{G}}, \omega_{\mathrm{L}} \sim \unit{0.1}{\mega\hertz}$. The Poissonian shot-to-shot fluctuations of the atomic number $N$, which can be approximated by a Gaussian as $N$ is large, result in fluctuations of $\Omega$ with a relative standard deviation $\sigma_{\Omega} \approx 0.024 \Omega$, increased to the experimentally-measured value $\sigma_\Omega\approx 0.028 \Omega$ by residual fluctuations of laser power. We account for these fluctuations by averaging the results of each simulation over 500 runs, drawing $\Omega$ from a Gaussian distribution.\\

We characterize these fluctuations by fitting the experimental Rabi oscillations in the regime $\Omega \gg \omega_{\mathrm{G}}, \omega_{\mathrm{L}}$, where the oscillation envelope is primarily governed by Rabi frequency fluctuations. Gaussian-distributed fluctuations in $\Omega$ are incorporated through the following fit function:
\begin{equation}
\label{eq:fluctRabiFreq}
f(t) = \frac{A}{2}\left(1 - \cos(\Omega t)\ee^{-\frac{1}{2}\sigma_{\Omega}^2 t^2}\right)\ee^{-\Gamma t},
\end{equation}
where $A$ is an amplitude scaling factor and $\Gamma$ the exponential decay rate represented in Fig.~6(e). Given the linear dependence of $\sigma_{\Omega}$ on $\Omega$, we fix the fluctuation parameter in the fit to extract $\Gamma$ for any Rabi frequencies. However, when the decay rate $\Gamma$ becomes small, due to to the protection mechanism, the envelope becomes dominated by fluctuations of $\Omega$. In this case, $\Gamma$ is difficult to estimate, leading to increased uncertainties for low values of $\Gamma$ visible in Fig.~6(e) of the main text.

\subsection{Inhomogeneous dephasing terms $\hat{H}_{\mathrm{Deph}}$ and $\mathcal{L}_{\mathrm{Deph}}$}

The Hamiltonian $\hat{H}_{\mathrm{Deph}}$  and the Liouvillian decay term $\mathcal{L}_{\mathrm{Deph}}$, describing inhomogeneous dephasing, are determined using the procedure described in the theoretical section of the manuscript and in the section~\ref{app:decay} above. We account for non-Markovianity by including in $\hat{H}_{\mathrm{Deph}}$  $n=15$ Dicke states for the purely Gaussian Doppler broadening, and $n=20$ for the light-shifted one.

For Gaussian dephasing, we extract $\omega_{\mathrm{G}} = 2\pi\times\unit{61(8)}{\kilo\hertz}$ from the fit of the measured population decay $|s_{\mathrm{G}}(t)|^2$ presented in ``Experimental results'' section of the main manuscript. This value is consistent with the temperature obtained by time-of-flight measurement with absorption imaging, yielding $\omega_{\mathrm{G}} = 2\pi\times\unit{66(1)}{\kilo\hertz}$.

\begin{figure}[tb]
\centering
\includegraphics[width=85mm]{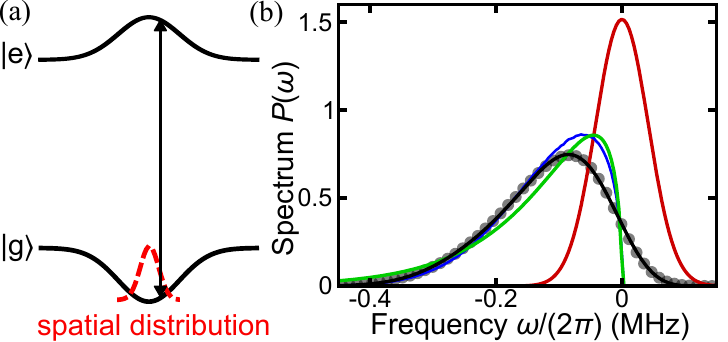}
\caption{ \textbf{Light-shift and Gaussian inhomogeneous dephasing spectra.} (a) Light shift induced by a Gaussian dipole trap. (b) Inhomogeneous dephasing spectra. Red: Gaussian experimental Doppler shift distribution. Green: light-shift distribution $ P_{\mathrm{L}}(\omega)$ for a thermal cloud, assuming that the trap is harmonic. Blue: light-shift distribution $ P_{\mathrm{L'}}(\omega)$ accounting for the trap's anharmonicity, obtained by Monte Carlo sampling using experimental parameters. Grey and black: full inhomogeneous dephasing spectrum combining a Gaussian Doppler broadening with a light shift in an anharmonic trap. Grey circles correspond to the Monte-Carlo-sampled spectrum, the full black line is a fit by $P_\mathrm{GL'}(\omega)=|\psi_{\mathrm{GL'},0}(\omega/(\sqrt{2}\omega_\mathrm{GL'})|^2/(\sqrt{2}\omega_\mathrm{GL'})$ where $\psi_{\mathrm{GL'},0}(x)$ is given by Eq.~\ref{eq:psiGLp}, the frequency scale $\omega_\mathrm{GL'}$ and the Fock state coefficients $\{\beta_k\}$ being used as free parameters.}
\label{fig:EneDis}
\end{figure}

For the light-shifted case, the scaling angular frequency $\omega_\mathrm{S}$ is obtained by independently measuring the temperature, the average light shift, and the width of the Gaussian atomic density profile in the trap. The anharmonicity of the trap, formed by crossed Gaussian beams, leads to a noticeable deviation between $\rho_\mathrm{L}$ and the actual light-shift distribution $\rho_\mathrm{L'}$, shown in Fig.~\ref{fig:EneDis}(b). We reproduce $\rho_\mathrm{L'}$ via Monte Carlo sampling, by randomly drawing a large number of atoms from the spatial distribution obtained by absorption imaging, calculating the light shift distribution knowing the trap's geometry, and rescaling it using the measured mean value of the  light shift. The velocity-dependent Doppler shifts being statistically independent on the position-dependent light shifts in a thermal cloud, the total inhomogeneous broadening distribution $\rho_\mathrm{GL'}$  is obtained either by a convolution of $\rho_\mathrm{L'}$ with a Gaussian $\rho_\mathrm{G}$ or by an additional Monte Carlo sampling of Gaussian-distributed atomic velocities. We center $\rho_\mathrm{GL'}$ and approximate the wavefunction of the symmetric Dicke state $\psi_0(x)=\sqrt{\rho_\mathrm{GL'}(x)}$ by a superposition of the first five Fock states,
\begin{equation}
\label{eq:psiGLp}
\psi_{\mathrm{GL'},0}(x) = \sum_{k=0}^4 \beta_k \frac{1}{\pi^{1/4}\sqrt{2^k k!}} \mathcal{H}_k(x)e^{-x^2/2},
\end{equation}
where $\beta_k$ are the coefficients and $\mathcal{H}_k$ are the Hermite polynomials. The procedure described in the theoretical section of the main manuscript is then applied algebraically, involving only the list of coefficients $\{\beta_k\}$.

\subsection{Additional loss term $\mathbf{\mathcal{L}_{\mathrm{Loss}}}$} 

Additional losses from the state $\ket{E_0}$, described by the Liouvillian term $\mathbf{\mathcal{L}_{\mathrm{Loss}}}$, arise from a combination of several effects left aside in the idealized model above. Among them, imperfections in the Rydberg blockade or a residual population of the short-lived intermediate state \ket{f} stemming from the finite detuning $\Delta$ are expected to induce loss rates scaling as $\Omega^2$, making the overall additional damping rate evolve as $\Gamma_{\Omega} \Omega^2 + \Gamma_0$. We estimate this rate in the strong driving regime with no light shifts, such as $\Omega \gg \omega_{\mathrm{G}}$, where driving protection makes inhomogeneous dephasing negligible. We fit the experimental oscillations to extract $\Gamma_{\Omega,\mathrm{fit}} \times 2 \pi = \unit{4.6(1) 10^{-3}}{\mega\hertz}^{-1}$ and $ \Gamma_{0,\mathrm{fit}} = \unit{2\pi \times 2.6(1.5)}{\kilo\hertz}$.

\end{document}